\documentclass[journal]{IEEEtran}

\ifCLASSINFOpdf
\else
   \usepackage[dvips]{graphicx}
\fi
\usepackage{url}
\usepackage[colorlinks,linkcolor=blue]{hyperref}
\usepackage{amsmath}
\usepackage{graphicx}
\usepackage{amssymb}
\usepackage{setspace}
\usepackage{caption}
\usepackage{multirow, threeparttable, booktabs, makecell}
\usepackage{notoccite}
\begin{document}

\title{Whisper-SV: Adapting Whisper for Low-data-resource Speaker Verification}
\author{Li Zhang$^{1,*}$, Ning Jiang$^{2,*}$ \thanks{* Co-first author.}, Qing Wang$^{1}$, Yue Li$^{1}$, Quan Lu${^2}$, and Lei Xie$^{1,\#}$\thanks{\# Corresponding author.}  \\ 
{$^1$Audio, Speech and Language Processing Group (ASLP@NPU), School of Computer Science, \\
Northwestern Polytechnical University (NPU), Xi'an, China}  \\
 { $^2$Mashang Consumer Finance Co.,Ltd, Beijing, China }  }

\maketitle
\begin{abstract}
Trained on 680,000 hours of massive speech data, Whisper is a multitasking, multilingual speech foundation model demonstrating superior performance in automatic speech recognition, translation, and language identification. However, its applicability in speaker verification (SV) tasks remains unexplored, particularly in low-data-resource scenarios where labeled speaker data in specific domains are limited.
To fill this gap, we propose a lightweight adaptor framework to boost SV with Whisper, namely Whisper-SV. Given that Whisper is not specifically optimized for SV tasks, we introduce a representation selection module to quantify the speaker-specific characteristics contained in each layer of Whisper and select the top-k layers with prominent discriminative speaker features. To aggregate pivotal speaker-related features while diminishing non-speaker redundancies across the selected top-k distinct layers of Whisper, we design a multi-layer aggregation module in Whisper-SV to integrate multi-layer representations into a singular, compacted representation for SV. In the multi-layer aggregation module, we employ convolutional layers with shortcut connections among different layers to refine speaker characteristics derived from multi-layer representations from Whisper. In addition, an attention aggregation layer is used to reduce non-speaker interference and amplify speaker-specific cues for SV tasks. Finally, a simple classification module is used for speaker classification. Experiments on VoxCeleb1, FFSVC, and IMSV datasets demonstrate that Whisper-SV achieves EER/minDCF of 2.22\%/0.307, 6.14\%/0.488, and 7.50\%/0.582, respectively, showing superior performance in low-data-resource SV scenarios.
\end{abstract}


\IEEEpeerreviewmaketitle
\vspace{-8pt}
\section{Introduction}

\IEEEPARstart{S}{peaker} verification~(SV) is the process of confirming if an individual is who they claim to be~\cite{naika2018overview}. In recent years, deep learning has shown remarkable success in SV tasks. However, current methods often rely on large amounts of labeled training speech~\cite{li2022cn,chung2018voxceleb2,desplanques2020ecapa,zhou2021resnext,xie2022global}, and the performance significantly declines in challenging scenarios with limited speaker labeled data, such as far-field and multi-language SV tasks.  
This decline is primarily due to the scarcity of large-scale speech datasets with speaker labels in these \textit{low-data-resource} scenarios~\cite{qin2020ffsvc,mishra2023msv} and the lack of robustness in speaker models trained on conventional Mel-Frequency Cepstral Coefficients (MFCC) and Filter Bank (Fbank) features. 

In general, data augmentation is commonly used in low-data-resource scenarios. Data augmentation includes various transformations of speech (e.g. add noise~\cite{snyder2015musan}, reverberation~\cite{habets2006room}, speed perturbation~\cite{yang2022data} and SpecAug~\cite{park2019specaugment}) and introducing out-of-domain datasets which may lead to domain mismatch problems~\cite{gusev2020stc,zhang2023distance}. 
In addition to the above data augmentation methods, leveraging pre-trained speech foundation models for downstream low-data-resource tasks has recently become increasingly popular due to their extensive exposure to massive amounts of data, enhancing their generalization capabilities for various tasks. Chen and Rudnicky \cite{chen2023exploring} explored wav2vec2.0 fine-tuning for improved low-data-resource speech emotion recognition. Gupta et al. \cite{gupta2023enhancing} enhanced language identification in low-data-resource languages (Indian) by exploiting learned features with wav2vec2.0. Zhao and Zhang \cite{zhao2022improving} exploited and analyzed a series of pre-trained models for speech recognition in 15 low-resource languages. Notably, using pre-trained large foundation speech models in downstream low-data-resource tasks facilitates the effective utilization of vast amounts of out-of-domain data to enhance model robustness and mitigate the domain mismatch issue typically associated with introducing out-of-domain data. Moreover, since these large foundation models are trained on extensive out-of-domain data, they often achieve good performance in downstream tasks without requiring a large number of trainable parameters.
  
Recently, a multi-task multi-lingual model called \textit{Whisper}~\cite{radford2023robust} was proposed, which was an encoder-decoder transformer architecture trained on a large-scale dataset of 680k hours of speech data, showing superior performances in various tasks, including multilingual automatic speech recognition (ASR), speech translation (ST), and language identification (LID). Although trained on vast data showing great potential in low-data-resource scenarios, Whisper is optimized for ASR, ST, and LID tasks, ignoring SV tasks that specialize in extracting and analyzing individual vocal characteristics for identity verification.
Therefore, we aim to explore its efficacy in SV, particularly focusing on low-data-resource SV scenarios where speaker-labeled data in the target domain is limited. 

 In this paper, we propose \textit{Whisper-SV}, a lightweight adaptor framework to transfer Whisper for SV tasks, particularly benefiting low-data-resource SV scenarios.
Given that Whisper is not explicitly designed for SV tasks, it is crucial to determine which layers of Whisper contain more discriminative speaker characteristics for SV, and how to effectively aggregate representations of multi-layer of Whisper for SV tasks. Consequently, Whisper-SV is elaborately designed to comprise four modules: a pre-trained Whisper module, a representation selection module, a multi-layer aggregation module, and a classification module, respectively.  Specifically, the pre-trained Whisper module is employed to provide robustness and generalized representations derived from Whisper pre-trained on diverse and massive speech datasets. The representation selection module quantifies the speaker-specific information within representations of each layer in Whisper and selects top-k layers containing crucial speaker-related characteristics. Subsequently, the multi-layer aggregation module is designed to aggregate pivotal speaker-related information
while reducing redundant features across the top-k distinct layers of Whisper. Finally, the classification module is tasked with speaker classification. 

The contributions of this work are as follows:
 \begin{itemize}
  \item We propose Whisper-SV, an adaptor framework for low-data-resource SV tasks, leveraging a small number of trainable parameters and a limited amount of speech data to transfer the robustness and generalization of representations from Whisper to low-data-resource SV.
  \item We introduce a representation selection module within Whisper-SV to quantify the speaker-specific information in representations of each layer in Whisper and select the top-k layers containing the most speaker-related information. Specifically, we train individual SV models for each layer of Whisper and evaluate their performance. 
  \item We design a multi-layer aggregation module in Whisper-SV to integrate multi-layer representations into a singular, compacted representation for SV. This module employs convolutional layers with shortcut connections among different layers to refine speaker characteristics derived from multi-layer representations. Additionally, an attention aggregation layer reduces non-speaker information and amplifies speaker-specific cues for SV tasks. 
\end{itemize}

 We extensively validate Whisper-SV through experiments conducted on low-data resource datasets~(VoxCeleb1 \cite{nagrani2020voxceleb}, FFSVC \cite{qin2020ffsvc}, and IMSV \cite{mishra2023msv}). The experimental results demonstrate that Whisper-SV achieves superior performance in low-data-resource SV scenarios despite having a small number of trainable parameters.
 
The remainder of the paper is structured as follows: Section II discusses related works, Section III outlines Whisper-SV, Section IV presents the experimental setup, Section V showcases results and analysis, Section VI concludes the paper, and Section VII explores future work.
 
\section{Related Work}
In this session, we discuss the related works of this paper from three perspectives: the application of Whisper in downstream tasks, the usage of pre-trained models in SV, and research on low-data-resource SV.
 
\subsection{Whisper for Downstream Task}
Whisper is trained on massive and diverse multilingual weakly supervised speech, specially tailored for multilingual ASR, ST, and LID tasks~\cite{radford2023robust}. The performance of Whisper achieves high-quality recognition results on various benchmarks without fine-tuning on specific datasets. 
Recently, some researchers have focused on adapting Whisper's generalization and robustness, which learned from massive amounts of data, to various downstream tasks.
Some studies delve into reducing the size of Whisper on streaming ST~\cite{machavcek2023turning} and efficient ASR~\cite{shao2023whisper}. Some strive to adapt Whisper for child speech recognition~\cite{jain2023adaptation,vasquez2023novel}. In addition to ASR, ST, and LID applications, some researchers are dedicated to transferring Whisper for other applications. Kodali et al.~\cite{kodali2023classification} investigate the embeddings of Whisper in the vocal intensity category. Wang et al.~\cite{wang2023can} introduce a speech-based in-context learning (SICL) approach with Whisper for test-time adaptation. Zezario et al.~\cite{zezario2023study} leverage the
acoustic features from Whisper for robust speech assessment and Rathod et al.~\cite{rathod2023noise} use the embeddings of Whisper for dysarthric severity-level classification. Ameer et al.~\cite{ameer2023whisper} enhance stuttered speech
classification by optimizing Whisper's encoder layer. WhisperSeg~\cite{gu2023positive} utilizes Whisper for human and animal voice activity detection, and Berns et al.~\cite{berns2023speaker} investigate Whisper on speaker and language change detection. Most of the above studies focus on utilizing representations of Whisper for downstream classification or detection tasks, which indicates the representations of Whisper encapsulate comprehensive speech features, encompassing acoustic characteristics, speaker attributes, vocal details, etc. Meanwhile, the generalization and robustness of Whisper in various speech tasks motivate us to explore its application in SV tasks. 
\vspace{-8pt}
\subsection{Pretrained Model for SV}
In the field of SV, pre-trained models usually fall into two categories, supervised pre-trained models and self-supervised learning (SSL) models respectively. Specifically, the supervised pre-trained models are trained with large-scale speaker-labeled datasets \cite{jung2022large,chung2018voxceleb2,li2022cn} and optimized for speaker classification tasks \cite{zhengspeakin}. The supervised pre-trained models tailored for SV are usually trained with large-scale speaker-labeled datasets (e.g. VoxCeleb~\cite{chung2018voxceleb2}, AISHELL-2~\cite{du2018aishell}, CNCELEB~\cite{fan2020cn}). The supervised pre-trained models designed for SV are commonly employed for initializing SV models~\cite{zhang2023distance,zhang2022npu,zhang2020npu} or applying domain adaptation techniques~\cite{zhang2021multi, lin2020multi,rohdin2019speaker} to transfer their robustness and generalization to low-data-resource SV scenarios. The SSL models are trained for general speech tasks (e.g. wavLM~\cite{chen2022wavlm}, Hubert~\cite{hsu2021hubert}, wav2vec~\cite{schneider2019wav2vec}). Many researchers leverage the weighted embeddings from all transformer layers and the conv-extractor module of SSL as features to train SV models~\cite{makarov2022id, huh2023voxsrc}. In addition, some studies focus on fine-tuning \cite{wang2021fine,novoselov2022robust} partial or entire SSL models for SV tasks, which introduce a large number of training parameters and extend the training duration. Whisper is a weak-supervised learning model for multi-lingual ASR, ST, and LID tasks. While some researchers have applied Whisper to various downstream tasks (e.g., speaker detection~\cite{berns2023speaker}, vocal intensity categorization~\cite{kodali2023classification}, dysarthric severity-level classification~\cite{rathod2023noise}, etc.), its potential in SV tasks remains unexplored and our paper aims to fill this gap.
\vspace{-8pt}

\begin{figure*}[th]
\centering
\centerline{\includegraphics[width=\linewidth]{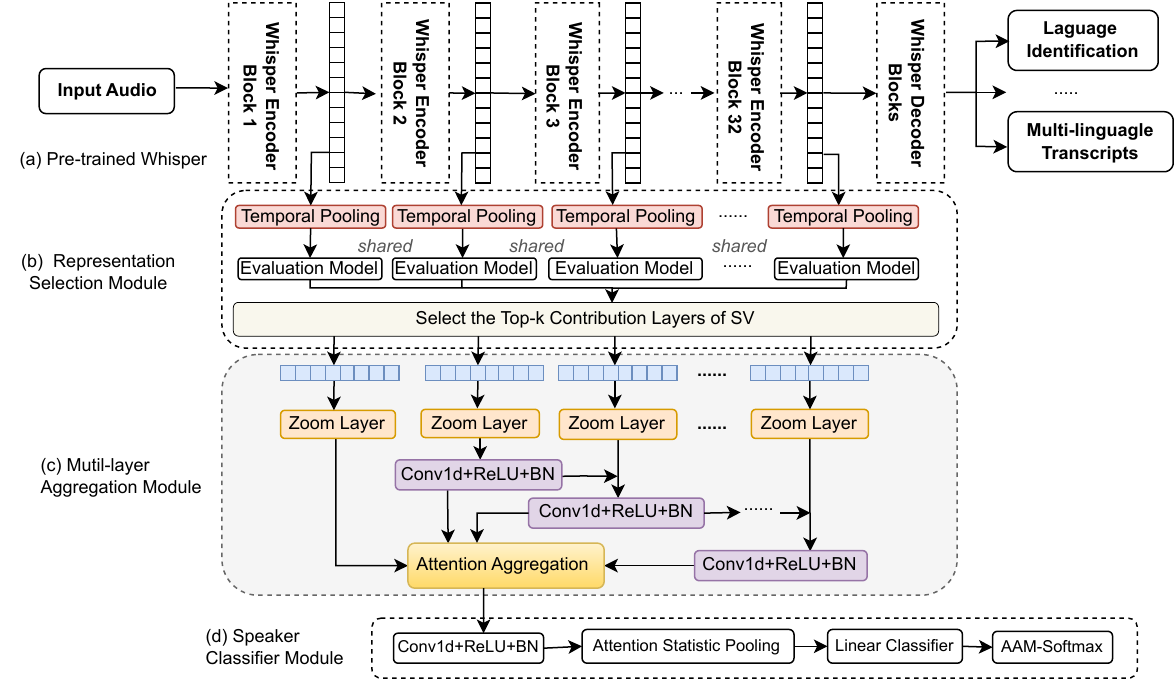}}
\caption{
The architecture of Whisper-SV includes four modules: (a) a pre-trained Whisper module for providing robust and generalized representations, (b) a representation selection module for selecting the top-k layers containing significant speaker-specific characteristics, (c) a multi-layer aggregation module to aggregate representations from multiple layers of Whisper, and (d) a speaker classifier module for speaker classification.}
\label{fig:Whisper-SV}
\vspace{-8pt}
\end{figure*}
\subsection{Low-data-resource SV}
Low-data-resource SV refers to developing SV systems with limited data resources, such as those involving minor languages and far-field SV applications. The limitation in data resources may arise from having a small number of speakers, insufficient diversity in speech content, or a lack of varied acoustic environments within the dataset.  The primary challenge lies in the scarcity of data that effectively captures the full spectrum of complexity and variability inherent in real-world speech, which is essential for training robust SV systems. To address the challenge of limited data availability, a straightforward approach is to augment the low-resource speech data. Conventional methods include adding noise~\cite{snyder2015musan} and reverberation~\cite{habets2006room}, speed perturbation~\cite{yang2022data}, and SpecAug~\cite{park2019specaugment}, among others. Some generative methods have also been employed to perform augmentation on low-data-resource SV~\cite{wang2021antvoice,zhang2022npu}. Besides synthesizing and augmenting low-resource speech, a large amount of out-of-domain data can also be utilized to improve speaker recognition performance in low-resource scenarios. Whether through synthetic data augmentation or the direct incorporation of out-of-domain speech data, the low-resource problem is transformed into a domain mismatch problem~\cite{rohdin2019speaker,lin2020multi,zhang2021multi}. Instead of introducing out-of-domain speech data, we explore the potential of the pre-trained large model Whisper to improve low-data-resource SV tasks.

\section{Methods}
The schematic diagram of Whisper-SV is depicted in Fig.~\ref{fig:Whisper-SV}, encompassing four modules: (a) a pre-trained Whisper module, (b) a representation selection module, (c) a multi-layer aggregation module, and (d) a speaker classifier module. The pre-trained Whisper module is to provide robust and generalized bottleneck representations from Whisper. Given that the principal training objective of Whisper is aligned with ASR, ST, and LID rather than SV tasks, the representation generated by the decoder of Whisper encompasses features that are more closely associated with ASR, ST, and LID. Consequently, in this paper, we focus on leveraging the representations produced by the encoder of Whisper to boost SV tasks. In addition, we incorporate a representation selection module for quantifying the speaker-specific information in representations of each layer in Whisper and select the top-k layers with prominent speaker-related discriminative features for SV. The multi-layer aggregation module is designed to integrate features across different layers of Whisper for SV. 
Finally, a simple classifier module is used for speaker classification. 

\subsection{Revisiting Whisper}
Whisper explores the potential of weakly supervised speech processing systems trained on diverse and extensive speech datasets. The evaluation results of Whisper have showcased the capacity of these systems to approach human-level performance when provided with sufficient training data~\cite{radford2023robust}. 
Whisper is based on a classical encoder-decoder Transformer~\cite{vaswani2017attention} architecture and the encoder is formed by two convolutional layers with a kernel size of 3, followed by a sinusoidal positional encoding, and a stacked set of
Transformer blocks. The decoder uses the learned positional
embeddings and the same number of Transformer blocks as the
encoder. Whisper is particularly optimized for ASR, ST, and LID tasks. Still, many studies delve into transferring the generalization and robustness of Whisper to other downstream tasks (e.g., vocal intensity classification~\cite{kodali2023classification}, dysarthric severity-level classification~\cite{rathod2023noise}, speaker activity detection~\cite{berns2023speaker, gu2023positive}, etc.). This paper focuses on adapting Whisper for SV tasks with a lightweight adaptor framework (Whisper-SV). In Whisper-SV, we use the encoder of the pre-trained Whisper as the bottleneck feature extractor for SV. As illustrated in Fig.~\ref{fig:Whisper-SV} (a), the encoder of Whisper transforms the input audio signals into a series of high-level feature representations.

Suppose we have a set of input audios, denoted as $X$. The corresponding speaker labels for $X$ are given by $Y = \{y_1, y_2, y_3, \ldots, y_i, \ldots, y_M\}$, and $M$ signifies the total number of speakers in the training set. The encoder blocks of Whisper are $F_W =\{f_w^1,f_w^2,f_w^3 ...f_w^l... f_w^L \}$, where $L$ is the layer number of Whisper encoder. After all input audio re-sampled to 16,000 Hz, we extract the multi-layer bottleneck representations from pre-trained Whisper as follows:
\begin{equation}
F_W(X) = \{f_w^1(X),f_w^2(X),f_w^3(X),...f_w^l(X)...,f_w^L(X) \},
\end{equation}
where $l \in \{1,2,3,..,L\}$ and $F_W(X)$ are the collections of representations of all layers in the Whisper encoder.

\subsection{Representation Selection Module}
Considering that Whisper is not explicitly optimized for SV tasks, the representation selection module is designed to evaluate speaker-specific characteristics within the representations of each layer in Whisper to identify the top-k layers with the most significant discriminative speaker information. The representation selection module comprises a temporal pooling layer $\rho$, evaluation models $F_S = \{ f_s^{1}, f_s^{2}, f_s^{3},...,f_s^{i},...,f_s^{L}\}$, and a strategy to select the top-k contribution layers for SV, as depicted in Fig. 1 (b). The temporal pooling layer $\rho$ is used to pool the bottleneck representations extracted from Whisper on the temporal dimension to the duration of the input audio. The evaluation models $F_S$, trained individually on the representations extracted from each layer of the Whisper encoder and subsequently evaluated on the validation dataset, are employed to quantitatively assess the speaker-related features encapsulated within the representation of each layer in Whisper. Suppose the assessment error rates of evaluation models are $P_S= \{P_s^{1},P_s^{2},...,P_s^{i},...,~P_s^{L}\}$ and $L$ is the layers of Whisper encoder. The error rate of the speaker model trained by the i-th layer representations is $P_s^{i} = \gamma(f_s^{i}(\rho(f_w^i(X)))) $, where $\gamma$ is the error metric of evaluation models $F_S$. $f_s^i$ is the i-th layer evaluation model. $\rho$ is the temporal pooling layer. The collection of error rates for speaker models trained individually on representations from each layer of Whisper is formulated as:

\begin{equation}
\begin{split}
P_S = \{ \gamma(f_s^{i}(\rho(f_w^i(X)))) \mid i = 1, 2, \ldots, L \} 
\label{eq2}
\end{split}
\end{equation}
In Eq.~(\ref{eq2}), we obtain the error rates of speaker models trained on the representations from each layer. We then identify the top-k layers by selecting those with the lowest error rates.
The selected $k$ layers of Whisper with the lowest error rate (the top-k layers containing the highest speaker-specific cues) are $P_S^{top-k}$:
\begin{equation}
\begin{aligned}
& P_S^{top-k} = \{P_s^{'1},P_s^{'2},P_s^{'3}, \ldots, P_s^{'k}\} \\
& \quad = \text{{argsort}}_{\text{{ascend}}}(P_S)[1:k].
\end{aligned}
\end{equation}
The representations associated with $P_S^{top-k}$ (contain crucial speaker-specific characteristics) in the top-k layers of Whisper are denoted as:
\begin{equation}
\tilde{F}_W^{top-k}(X)=\{\tilde{F}_w^1(X), \tilde{F}_w^2(X), \tilde{F}_w^3(X), \ldots, \tilde{F}_w^k(X)\}.
\end{equation}
\subsection{Multi-layer Aggregation Module}
To aggregate multi-layer representations from Whisper into a single, condensed representation for SV, we introduce a multi-layer aggregation module to amalgamate the representations from the top-k layers in Whisper for SV. As illustrated in Fig. 1 (c), the multi-layer aggregation module consists of zoom layers, multi-level aggregation layers based on 1D convolution, and an attention aggregation layer for fusing speaker-specific cues and suppressing speaker-irrelevant information from representations of top-k layers in Whisper.

The zoom layers  $F_z =\{F_z^{1},F_z^{2},F_z^{3},...,F_z^{l},F_z^{k}\}$
are used to scale the dimensions of representations extracted from the Whisper encoder. The zoom layers are essentially one-dimensional convolutional layers engineered to modulate the channel dimension to a reduced scale. The primary function of these zoom layers is to attenuate extraneous information about non-speakers while concurrently extracting salient discriminative features for SV tasks. This process not only enhances the model's specificity but also contributes to a reduction in the overall parameter footprint of SV models. The outputs of the zoom layers are:
\begin{equation}
F_z^{top-k}(X) =\{F_z^{1}(\tilde{F}_w^1(X)),F_z^2(\tilde{F}_w^2(X)),...,F_z^k(\tilde{F}_w^k(X))\}.
\end{equation}

Subsequently, the multi-level aggregation layers based on 1D convolution are introduced to aggregate speaker-specific cues across representations of top-k layers in the Whisper encoder. Besides the top-1 representation of $F_z^{1}(\tilde{F}_w^1(X))$, the other scaled representations are individually fed into separate convolution blocks, and each convolution block consisting of one convolution layer, one ReLU layer, and one batch normalization layer. There are $k-1$ convolution blocks representing as $F_M =\{F_m^{2},...,F_m^{k}\}$. To better aggregate information across multiple layers, shortcut links are used among convolution blocks, where the input to the convolution block of the $i$-th layer is the sum of the output from the convolution block of the $(i-1)$-th layer and the output from the zoom layer of the $i$-th layer. The input of the $i$-th convolutional block is:
\begin{equation}
 F_m^i(X) = F_m^{(i-1)}(F_z^{i-1}(\tilde{F}_w^{i-1}(X))) \oplus  F_z^{i}(\tilde{F}_w^i(X)),
\end{equation}
where $2 < i \leq k$ and $\oplus$ represents the short-cut connection.
The outputs of the $k-1$ convolution modules and the scaling layer of the top-1 layer together form the input to the attention aggregation layer.
\begin{equation}
 F_a(X) = \frac{1}{T}(cat[F_z^{1}(X),F_m^2(X),...F_m^k(X)]),
\end{equation}
where $F_a(X) \in \mathbb{R}^{B \times C \times T}$ and $B$, $C$, $T$ are the batch size, channels, and frame numbers of the aggregated representation.

In the attention aggregation layer, we employ both squeeze-and-excitation (SE) attention~\cite{hu2018squeeze} and self-attention~\cite{vaswani2017attention} to aggregate representations from different layers of the whisper encoder for SV tasks. However, the results show that SE attention performs better, possibly because SE attention recalibrates feature channels, reinforcing the learning of important features while suppressing unimportant ones. Different channels may contain different semantic information, and SE attention can effectively distinguish the importance of speaker-related characteristics. The formulation of the SE attention is expressed as:
\begin{equation}
M^C = \delta (W_2^{C \times C'} * (relu((W_1^{C' \times C} * F_a(X))))),
\label{SE_Formula}
\end{equation}
where $*$ means matrix multiplication. $W_1^{C' \times C}$ and $W_2^{C \times C'}$ are two fully connected layers that capture the inter-dependence of channels in the aggregated representation $F_a(X)$. The dimensional reduction factor $\frac{C}{C'}$ indicates the aggregated representation reduction ratio to avoid the parameters overhead. Finally, a sigmoid function $\delta$ scales the channel-wise weights. The attention mask is a set of attention factors predicted by supervised speaker classification loss aiming to emphasize essential speaker-specific features and compress the non-speaker information from Whisper. The output of the multi-layer aggregation layer is:
\begin{equation}
 F'_a (X) = M^C  \cdot F_a(X),
\end{equation}
where $\cdot$ is element-point multiplication.

\subsection{Speaker Classifier Module}
As illustrated in Fig. 1 (d), the speaker classifier module consists of a convolutional layer, an attention statistical pooling (ASP) layer~\cite{desplanques2020ecapa}, a linear classification layer, and AAM-softmax~\cite{deng2019arcface} as the loss function. Suppose the parameters of the speaker classifier module are $F_c$, and the final prediction speaker labels are $F_c(F'_a(X))$. The loss is formulated as:
\begin{align}
\begin{split}
& \mathcal{L}(X, Y) = \mathop{\mathbb{E}}_{(X,Y) \sim \mathcal{P}}[F_c(F'_a(X)), Y)],
\end{split}
\end{align}
where $F_c$ denotes all layers of the speaker classifier module. 
\subsection{Parameters Analysis}

In Whisper-SV, the parameter size of each module during the inference and training stages is shown in TABLE \ref{parameters}. The Representation Selection Module is used before training to evaluate which layers of Whisper's embeddings have more distinctive speaker-related information, and it does not participate in training and inference. The multi-layer aggregation module and the speaker classifier module undergo joint training. The pre-trained Whisper module is a feature extractor and is not involved in the training process. To conserve time and computational resources, we extract data representations using the pre-trained Whisper model once and refrain from reutilizing it throughout the iteration process.
\begin{table}[th]
\caption{Parameters of Whisper-SV in Inference and Training Stage}
\label{parameters}
\resizebox{\linewidth}{!}{\begin{tabular}{lcc}
\toprule
\hline
\multicolumn{1}{c}{Module}          & \begin{tabular}[c]{@{}c@{}}Inference \\ Parameters (M)\end{tabular} & \begin{tabular}[c]{@{}c@{}}Trainable \\ Parameters (M)\end{tabular} \\ \hline
(a) Pre-trained Whisper             & 1550                                                                & 0                                                                   \\
(b) Representation Selection Module & 0                                                                   & 0                                                                   \\
(c) Multi-layer Aggregation Module  & 4.610                                                               & 4.610                                                               \\
(d) Speaker Classifier Module       & 0.725                                                               & 0.725                                                               \\ \hline
\end{tabular}}
\end{table}

\section{Experimental Setup}
\label{sec:guidelines}

\subsection{Dataset}
In this paper, we evaluate the performance of Whisper-SV on  VoxCeleb1~\cite{nagrani2020voxceleb}, FFSVC~\cite{qin2020ffsvc}, and IMSV~\cite{mishra2023msv} datasets. VoxCeleb1 is collected from open-source media. The VoxCeleb1 training set comprises 1211 speakers, with a total duration of approximately 340 hours. FFSVC is a multi-channel far-field speaker recognition dataset. In this paper, we utilize the speech recordings from the second channel and iPhone in both the training and test sets of FFSVC (referred to as 'FFSVC' hereafter) for conducting experiments. Specifically, the FFSVC dataset comprises 120 speakers and approximately 288 hours of audio. IMSV is an indic-multilingual speaker recognition dataset. The IMSV training set comprises 50 speakers, with a total duration of approximately 100 hours. The test set we use includes VoxCeleb-O~\cite{nagrani2020voxceleb}, FFSVC2020 Task1 development trials (the second channel speech for testing)~\cite{qin2020ffsvc}, and IMSV development trials~\cite{mishra2023msv}. The data volume of VoxCeleb1, FFSVC, and IMSV is comparatively limited, rendering them suitable as low-resource datasets for SV tasks. 
Furthermore, to assess Whisper-SV's performance in scenarios with severe data limitations, we conduct experiments by randomly partitioning training datasets from VoxCeleb1, FFSVC, and IMSV into fractions of the original data volume—specifically, whole, half (1/2), quarter (1/4), and one\_eighth (1/8) while ensuring a consistent number of speakers in the training process.

In addition, we compare the results of Whisper-SV with data augmentation methods. Data augmentation is performed on Musan~\cite{snyder2015musan} and room impulse response~\cite{habets2006room} datasets. It's worth noting that we do not apply any data augmentation during the training process of Whisper-SV.

\subsection{Configurations}
\textbf{Model Configurations}: 
In this paper, we employ the Whisper-large-v2~\cite{radford2023robust} as the pre-trained Whisper module. The evaluation model in the representation selection module is ECAPA-TDNN with configurations of [1024, 1024, 1024, 3072]~\cite{desplanques2020ecapa}. Ablation experiments indicate that Whisper-SV performs optimally with a top-k value of 4 while having a relatively modest number of parameters. The configurations of components of the multi-layer aggregation module and speaker classification module are illustrated in TABLE 
\ref{table:configurations_model}.
\begin{table}[th]
\caption{The configurations of Whisper-SV.}
\label{table:configurations_model}
\resizebox{\linewidth}{!}{ \begin{tabular}{cccc}
\toprule
\hline
\multicolumn{1}{c}{Layer}           & \multicolumn{1}{c}{Channels} & \multicolumn{1}{c}{Kernel} & \multicolumn{1}{c}{Dilation} \\ \hline
Zoom Layer                          & 128                          & 5                          & 1                            \\
Conv1d in Multi-layer Aggregation Module                & 512                         & 3                          & 3                            \\
Attention Aggregation Layer          & 128                          & -                          & -                            \\
Conv1d in Speaker Classifier Module & 256                          & 1                          & 1                            \\
Attention Statistic Pooling Layer   & 128                          & -                          & -                            \\ \hline
 \end{tabular}}
\end{table}
 
\textbf{Training Configurations}: 
In this paper, we employ two types of feature extraction methods. The first method is applied to the input features of Whisper, which computes 80-dimensional log magnitude Mel spectrogram features using 25-millisecond windows and a 10-millisecond stride. The second method utilizes 80-dimensional Filterbank (Fbank) features with a 25ms window size and a 10ms frameshift, serving as the baseline for training the comparison models. Before feature extraction, the input audio is resampled at 16,000 Hz. The loss function employed is the additive angular margin softmax~(AAM-softmax)~\cite{deng2019arcface} with a margin parameter set to 0.2 and a scale parameter set to 30. Additionally, a weight decay of 2e-5 is applied in the training process. The optimization employs the Adam optimizer with a cyclic learning rate, which varies between 1e-8 and 1e-3, following the triangular policy~\cite{smith2017cyclical}.

\subsection{Comparison Methods}
We conduct a comparative analysis of Whisper-SV and other competitive methods. These methods include ECAPA-TDNN trained with Fbank~(ECAPA-TDNN~(Fbank)) and representations (the selected top-1 representations by the representation selection module) of Whisper~(ECAPA-TDNN~(Whisper)), data augmentation, supervised pre-trained SV models, SSL pre-trained models, and some domain adaptation methods in low-data-resource SV. The comparison methods are listed in TABLE~\ref{tab:Comparison }.
\begin{table}[th]
\caption{Comparison Methods }
\label{tab:Comparison }
\resizebox{\linewidth}{!}{ \begin{tabular}{cl}
\toprule
\hline
Category                                  & \multicolumn{1}{c}{Methods}                                                                                             \\ \hline
\multirow{2}{*}{Baseline}                 &  ECAPA-TDNN (Fbank)~\cite{desplanques2020ecapa}                                                                                      \\
                                          &  ECAPA-TDNN (Whisper)                                                                         \\ \cline{2-2} 
\multirow{2}{*}{Data Augmentation}        & \begin{tabular}[c]{@{}l@{}}Data augmentation (add noise~\cite{snyder2018x} and reverberation~\cite{snyder2018x},  \\ SpecAug~\cite{park2019specaugment}, speedPerturb~\cite{yang2022data}) \end{tabular} \\
                                          \cline{2-2} 
\multirow{2}{*}{SSL Method}               & Entirely finetuning Hubert for SV~(EF-Hubert-large-finetune)~\cite{wang2021fine}                                                                                                 \\
                                          & Entirely finetuning wav2vec for SV~(EF-wav2vec-large-finetune)~\cite{wang2021fine}                                                                                                  \\ \cline{2-2} 
\multirow{3}{*}{Domain Adaptation} & Wasserterin domain adversarial~\cite{rohdin2019speaker} training                                                                               \\
                                          & MMD tranfer learning~\cite{lin2020multi}                                                                                                   \\
                                          & WTR finetuning~\cite{zhang2023distance}                                                                                                          \\ \cline{2-2} 
\multirow{5}{*}{
 Model and Loss Improvements}      & Dual-model self-regularization and fusion~(DualA-tFC)~\cite{duan2023dual}                                                           \\
                                        
                                          & Angular margin prototypical loss (AMP Loss)~\cite{thanh2021deep}                                                                                       \\
                                          & ConvNext-Tiny~\cite{kizitskyi2023improving}                                                                                                          \\
                                          & Lighten CNN-LSMT~\cite{zhao2019lighten}                                                                                                        \\ \hline
\end{tabular}
}
\end{table}
\vspace{-6pt}
\subsection{Metric Score}
In the test phase, we use cosine similarity as the scoring criterion. The performance metrics are equal error rate (EER)
and minimum detection cost function (minDCF)~\cite{martin2000nist} which is
evaluated with $P_{target} = 0.01$, $C_{miss} = C_{fa} = 1$. Furthermore, we incorporate two critical metrics: the number of parameters (measured in millions, M) and the floating-point operations per second (FLOPs, quantified in giga, G). In this paper, all FLOPs calculations are performed with a batch size of 1.

\section{Experimental Results}

\subsection{Analysis of Representations of Whisper}
\begin{figure}[th]
\centering
\centerline{\includegraphics[width=\linewidth,height=14cm]{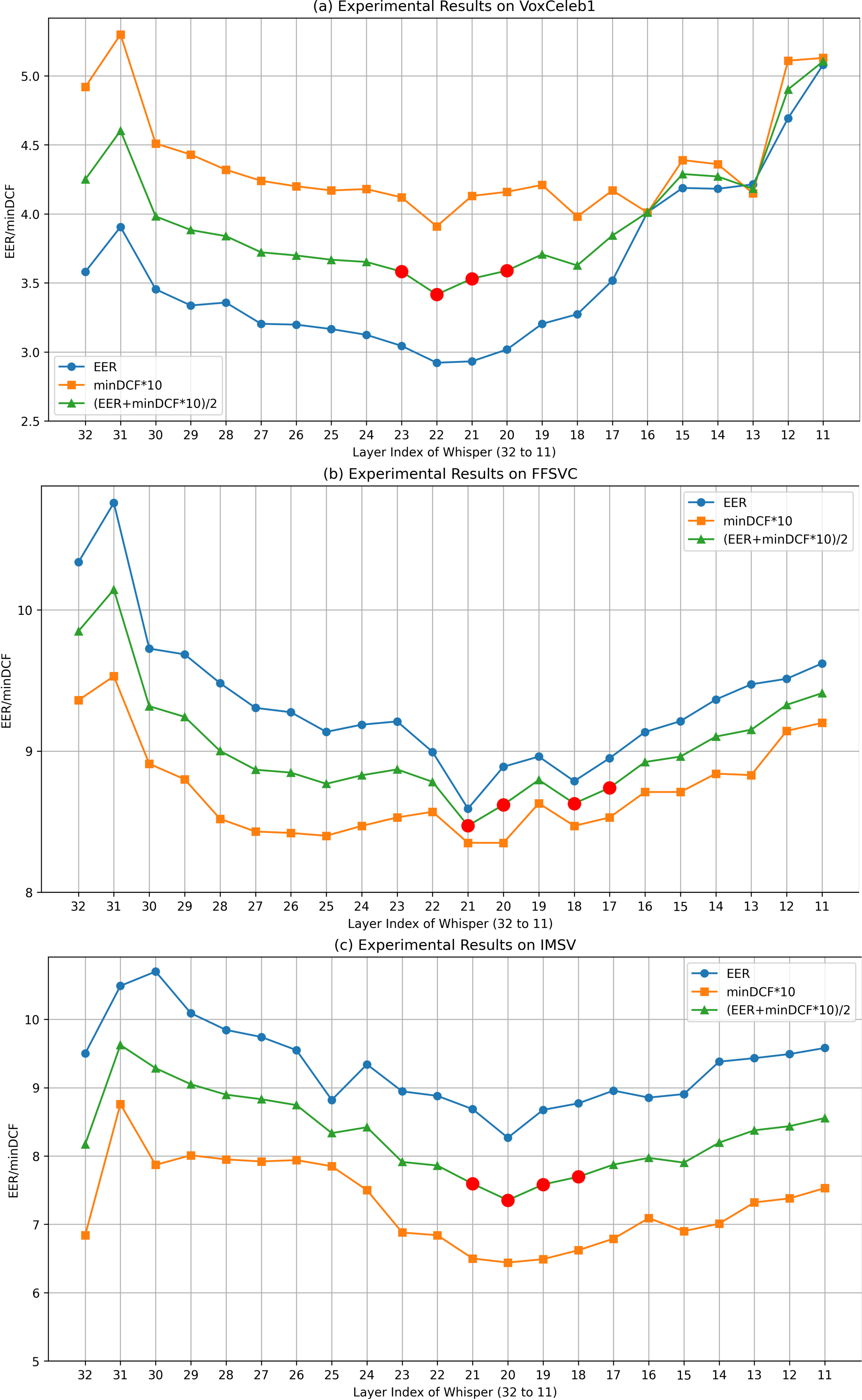}}
\caption{Experimental results of ECAPA-TDNN trained with representations extracted from each layer of Whisper. The red dots indicate the top four lowest combined EER and minDCF ((EER +10*minDCF)/2).}
\label{fig:results_layers}
\end{figure}

Considering that Whisper is purpose-built for multi-lingual  ASR, ST, and LID, to tailor Whisper for SV tasks effectively, it is crucial to quantify speaker-specific cues inherent within representations of each layer in Whisper. 
To thoroughly analyze the speaker-related discriminative information encapsulated within representations of each distinct layer, we independently train the SV model utilizing representations extracted from each layer of Whisper and assess the performance of each layer's representations in SV. The experimental results are depicted in Fig.~\ref{fig:results_layers}. We employ a composite evaluation metric ($(EER+10*minDCF)/2$) to identify the top-k layers with the lowest error rate to determine the optimal representation extraction layers for training SV models in Whisper. This approach provides a comprehensive assessment by jointly considering the EER and minDCF metrics, as illustrated by the green lines in Fig. \ref{fig:results_layers}. 

 Fig. \ref{fig:results_layers} displays experimental results from three datasets (VoxCeleb1, FFSVC, and IMSV), indicating that Whisper's representations around the 20th layer exhibit optimal speaker discriminative characteristics for SV tasks. Furthermore, the trend observed in the three sub-graphs suggests diminishing speaker-specific information of Whisper representations before layer 15. Consequently,  our experiments thoroughly analyze the performance of representations from layer 32 to layer 11 of Whisper in SV tasks. In  Fig.~\ref{fig:results_layers} (a), (b) and (c), the four red dots represent the top-4 layers with significant speaker discriminative characteristics for the corresponding datasets. The results of the selected top-4 layers are illustrated in TABLE \ref{four_best_results}. As evident from TABLE \ref{four_best_results}, the layers containing the most crucial speaker-discriminative information within the Whisper framework are predominantly centered around the 20th layer, located in the middle rear of the Whisper encoder. 
 Based on these experimental findings, we are going to identify and harness the top-k layers that demonstrate optimal performance impacts on the SV model. These selected layers are integrated within a multi-layer aggregation module specifically designed for boosting SV tasks.

\begin{table}[th]
\caption{EER (\%) and minDCF (p=0.01) of ECAPA-TDNN trained with representations of top-4 layers of Whisper.}
\label{four_best_results}
\resizebox{\linewidth}{!}{\begin{tabular}{ccccc}
\toprule
\hline
Training Dataset                &  Layer Index & EER    & minDCF & (EER+minDCF*10)/2 \\ \hline
\multirow{4}{*}{VoxCeleb1} & 22          & \textbf{2.92}  & \textbf{0.391}  & \textbf{3.415}             \\
                          & 21          & 2.94  & 0.413  & 3.535             \\
                          & 20          & 3.02  & 0.416  & 3.590            \\
                          & 23          & 3.05  & 0.418  & 3.615             \\ \cline{2-5} 
\multirow{4}{*}{FFSVC}    & 21          & \textbf{8.59 }  & \textbf{0.835}  & \textbf{ 8.470}            \\
 & 18          & 8.79          & 0.847              & 8.630   \\
                          & 20          & 8.89    & 0.845               & 8.670           \\
                                
                          & 17          & 8.96     & 0.853  & 8.745            \\ \cline{2-5} 
\multirow{4}{*}{IMSV (1/4)}     & 20          & 
\textbf{8.50}  & \textbf{0.649}  & \textbf{7.495} \\
                          & 19          & 8.67  & 0.649  & 7.580            \\
                          & 21          & 8.68  & 0.650  & 7.590            \\
                          & 18          & 8.77  & 0.662  & 7.695            \\ \hline
\end{tabular}}
\end{table}
\vspace{-5pt}
 
\subsection{Experimental Results on VoxCeleb1}
The experimental results on VoxCeleb1 are presented in TABLE \ref{voxceleb1_results}.  TABLE \ref{voxceleb1_results} highlights that using raw representations extracted directly from Whisper (Raw Whisper Representation) for SV is unviable, yielding an EER of 44.37\% and a minDCF of 0.999. This challenge primarily stems from Whisper not being optimized explicitly for SV tasks. 
 Therefore, it is essential to adapt Whisper for SV through the development of Whisper-SV.

When compared to training ECAPA-TDNN directly using the optimal representations of the top-1 layer in Whisper (ECAPA-TDNN~(Whisper)), Whisper-SV demonstrates a relative improvement of 23.9\%/21.4\% on EER/minDCF. Meanwhile, the outcomes of ECAPA-TDNN~(Whisper) surpass those achieved by ECAPA-TDNN trained with Frank (ECAPA-TDNN~(Fbank)). This observation underscores that representations extracted from Whisper encompass richer discriminative speaker-specific features SV tasks.
In addition, when compared with fine-tuning self-supervised models (wav2vec2.0~\cite{schneider2019wav2vec} and Hubert~\cite{hsu2021hubert}) for SV, Whisper-SV achieves lower EER and minDCF with a smaller number of trainable parameters and computational efficiency. Furthermore, Whisper-SV outperforms the Siamese capsule back-end processing method and the binary networks in SV tasks.
 
\begin{table}[th]
\caption{Experimental results on VoxCeleb (EER(\%), minDCF (p=0.01)).}
\label{voxceleb1_results}
\resizebox{\linewidth}{!}{\begin{tabular}{lccccc}
\toprule
\hline  
\multirow{2}{*}{Method}            & \multirow{2}{*}{\begin{tabular}[c]{@{}c@{}}Training\\  Dataset\end{tabular}} & \multirow{2}{*}{\begin{tabular}[c]{@{}c@{}}Trainable\\ Parameters (M)\end{tabular}} & \multirow{2}{*}{Flops (G)} & \multicolumn{2}{c}{VoxCeleb-O} \\ \cline{5-6} 
                                   &                                                                              &                                                                                     &                            & EER           & minDCF         \\ \hline
ECAPA-TDNN (Fbank)~\cite{desplanques2020ecapa}                   & VoxCeleb1                                                                    & 20.768                                                                              & 9.6                        & 4.37          & 0.379          \\
w/ Data augmentation                & VoxCeleb1                                                                    & 20.768                                                                              & 9.6                        & 3.02          & 0.326          \\
Raw Whisper Representation         & VoxCeleb1                                                                    & 0                                                                                   & 9.6                        & 44.37         & 0.999          \\
ECAPA-TDNN (Whisper)               & VoxCeleb1                                                                    & 20.768                                                                              & 9.6                        & 2.92          & 0.391          \\
EF-Hubert-large-finetune~\cite{wang2021fine}           & VoxCeleb1                                                                    & 300                                                                                 & 9.6                        & 2.36          & -              \\
EF-wav2vec-large-finetune~\cite{wang2021fine}          & VoxCeleb1                                                                    & 317                                                                                 & -                          & 3.42          & -              \\
Siamese Capsule~\cite{hajavi2021siamese}                    & VoxCeleb1                                                                    & 8.3                                                                                 & -                          & 3.14          & -              \\

Linear Weighted Sum~-~32 Layers      & VoxCeleb1                                                                    & 1.215                                                                               & 0.296                      & 3.26          & 0.369          \\
Random Four Layer~-~Layer~[22,21,1,8] & VoxCeleb1                                                                    & 5.336                                                                               & 0.643                      & 2.97          & 0.417          \\
Average Pooling~Top-4 Layers      & VoxCeleb1                                                                    & 2.597                                                                               & 0.581                      & 2.91          & 0.330               \\
Whisper-SV                         & VoxCeleb1                                                                    & 5.336                                                                               & 0.643                      &\textbf{ 2.22}          & \textbf{0.307}   \\
w/o Zoom Layer (No Zoom Layer)                      & VoxCeleb1                                                                    & 21.647                                                                              & 5.458                      & 2.77          & 0.428            \\
    
\hline
\end{tabular}}
\end{table}

Additionally, we explore alternative methods for using Whisper representations, including 1)~the commonly used linear weighted sum of representations from all 32 layers (Linear Weighted Sum~-~32 Layers), 2)~training Whisper-SV with representations from randomly selected four layers (Random Four Layers~-~Layer~[22, 21, 1, 8]), 3)~excluding the use of the zoom layer (i.e., not performing dimensionality reduction on Whisper representations) (w/o Zoom Layer), and 4)~using average pooling of representations from the top-4 layers instead of the multi-layer aggregation module (Average Pooling~-~Top-4 Layers). As the experimental results shown in TABLE~\ref{voxceleb1_results} indicate, the performance of these alternative approaches has not surpassed that of Whisper-SV. The analysis of these alternative methods is as follows. The Linear Weighted Sum of 32 layers may introduce non-speaker-related information, as Whisper was initially trained for ASR tasks, not SV tasks. Not all layer representations are equally suitable for SV tasks. Randomly selecting four layers may degrade SV performance if shallow representations are chosen, as illustrated in Fig.~\ref{fig:results_layers}, where shallow representations perform poorly for SV tasks. Removing the zoom layer significantly increases the number of parameters, as Whisper representations have a dimensionality of 1280. It may also lead to overfitting, as the zoom layer is primarily used for dimensionality reduction and extracting speaker-related information from Whisper representations. The use of average pooling of representations from the top-k layers, instead of the original multi-layer aggregation, does not purify and fuse representations from different layers, introducing non-speaker-related information. Average pooling weakens the discriminative power of speaker-related solid representations. 
\subsection{Experimental Results on FFSVC}
The experimental results for FFSVC are presented in TABLE \ref{FFSVC_results}. Notably, the experimental results of Whisper-SV outperform those of ECAPA-TDNN (Fbank) and ECAPA-TDNN (Whisper). Furthermore, we discover that utilizing only 1/4 of the FFSVC training dataset yields superior results compared to using the entire dataset. This can be attributed to Whisper's robust denoising and discriminative feature extraction capabilities, and make Whisper-SV necessitate fewer speech samples from a single speaker. Excessive data of per speaker may result in overfitting. Therefore, Whisper-SV is well-suited for data low-resource SV scenarios.

Additionally, we conduct comparison experiments between Whisper-SV and domain adaptation methods (vanilla fine-tuning, WTR finetuning, and Wasserterin DA), which are trained with VoxCeleb2~\cite{chung2018voxceleb2} and FFSVC dataset. The results indicate that Whisper-SV trained with only 1/4 of the FFSVC data outperforms vanilla fine-tuning and Wasserterin DA methods. Utilizing a vanilla fine-tuned Whisper-SV model pre-trained on VoxCeleb1, we observe a notable 5.9\% decrease in minDCF relative to the minDCF achieved by the WTR method, despite the WTR method exhibiting a lower EER. Meanwhile, Whisper-SV maintains a relatively low number of trainable parameters and FLOPs. These findings demonstrate that Whisper-SV can perform better with lower data requirements and fewer trainable parameters.

\begin{table}[th]
\caption{Experimental results on FFSVC (EER(\%), minDCF (p=0.01)).}
\label{FFSVC_results}
\resizebox{\linewidth}{!}{\begin{tabular}{lccccc}
\toprule
\hline
\multirow{3}{*}{Method} & \multirow{3}{*}{Training Dataset} & \multirow{3}{*}{\begin{tabular}[c]{@{}c}Trainable\\ Parameters (M)\end{tabular}}  & \multirow{3}{*}{FLOPs~(G)} & \multicolumn{2}{c}{FFSVC} \\
                        &   \cline{4-5}                              &                             &                        &    & EER   &  minDCF\\ \cline{1-6}  ECAPA-TDNN (Fbank)~\cite{desplanques2020ecapa}       & FFSVC                               & 20.768                       & 4.814                   & 9.49                                           & 0.810                                           \\
\quad + Data augmentation      & FFSVC                              & 20.768                       & 4.814                   & 8.55                                           & 0.774                                           \\
ECAPA-TDNN (Whisper)     & FFSVC                              & 20.768                       & 4.814                   & 8.59                                          &0.835                                         \\
ECAPA-TDNN (Whisper)     & FFSVC (1/4)                        & 20.768                       & 4.814                   & 8.56                                          & 0.828                                        \\
Vanilla Finetuning               & VoxCeleb2 + FFSVC                  & 20.768                       & 4.814                   & 8.44                                           & 0.548                                           \\
WTR~\cite{zhang2023distance}                    & VoxCeleb2 + FFSVC                  & 20.768                       & 4.814                   & \textbf{5.53} & 0.519                                           \\
Wasserterin DA~\cite{liu2020text}           & VoxCeleb2 + FFSVC                  & 20.769                       & 4.815                   & 5.78                                          & 0.5697                                          \\
Whisper-SV               & FFSVC (1/4)                        & 5.336                        & 0.643                   & 7.90                                           & 0.631                                           \\
Whisper-SV Finetuning    & VoxCeleb1 + FFSVC (1/4)            & \textbf{5.336}                        & \textbf{0.643}                   & 6.14                                           &  \textbf{0.488}     \\ \hline
\end{tabular}
}
\end{table}
\vspace{-12pt}
\subsection{Experimental Results on IMSV}

The experimental results for IMSV are displayed in TABLE \ref{IMSV_results}. As demonstrated in TABLE \ref{FFSVC_results}, the IMSV results also reveal that using only 1/4 of the training data leads to lower EER results than utilizing the entire dataset. Compared with ECAPA-TDNN (Whisper), Whisper-SV shows a relative reduction of 11.7\%.

\begin{table}[th]
\caption{Experimental results on IMSV (EER(\%), minDCF (p=0.01)).}
\label{IMSV_results}
\resizebox{\linewidth}{!}{\begin{tabular}{lccccc}
\toprule
\hline
\multirow{3}{*}{Method} & \multirow{3}{*}{Training Dataset} &  \multirow{3}{*}{\begin{tabular}[c]{@{}c}Trainable \\ Parameters (M)\end{tabular}}  & \multirow{3}{*}{FLOPs~(G)} & \multicolumn{2}{c}{IMSV} \\
                        &   \cline{4-5}                              &                             &                        &    & EER   &  minDCF\\ \cline{1-6} 
ECAPA-TDNN (Fbank)~\cite{desplanques2020ecapa}      & IMSV (1/4)                        & 20.768                      & 4.814                  & 11.51  & 0.845          \\
\quad + Data augmentation     & IMSV (1/4)                         & 20.768                      & 4.814                  & 9.23  & 0.845           \\
ECAPA-TDNN (Whisper)    & IMSV                               & 20.768                      & 4.814                  & 8.90  & 0.604         \\
ECAPA-TDNN (Whisper)    & IMSV (1/4)                         & 20.768                      & 4.814                  & 8.50  & 0.649          \\
ConvNext-Tiny~\cite{kizitskyi2023improving}           & IMSV                               & 27.796                      & 2.903                  & 19.01 & 0.987           \\
AMP Loss~\cite{thanh2021deep}                & IMSV                               & 20.768                      & 4.814                  & 7.97  & 0.827           \\

CNN-LSTM~\cite{zhao2019lighten}                & IMSV                               & 0.587                       & 0.025                  & 43.86  & 0.987           \\
Whisper-SV              & IMSV (1/4)                          & 5.336                       & 0.643                  & \textbf{7.50}  & \textbf{0.582 }         \\ \hline
\end{tabular}}
\end{table}

\subsection{Extension Experiments on VoxCeleb2}
To further validate the effectiveness of Whisper-SV on general SV tasks, we conduct experiments on the VoxCeleb2 dataset as shown in TABLE \ref{voxceleb2_results}. When VoxCeleb2 is used as the training set, Whisper-SV achieves EER/minDCF of 1.71\%/0.211 on the VoxCeleb-O. This performance, compared to the first-place result in VoxSRC2019 using ResNet34 with 256 channels \cite{chung2019voxsrc,zeinali2019but}, exhibits a slight increase in EER by 0.29\%. However, it surpasses the results obtained with the X-vector model \cite{wang2021revisiting,snyder2018x} of equivalent parameter sizes.
\begin{table}[th]
\caption{Experimental results on VoxCeleb2 (EER(\%), minDCF (p=0.01)).}
\label{voxceleb2_results}
\resizebox{\linewidth}{!}{\begin{tabular}{lccccc}
\toprule
\hline  
\multirow{2}{*}{Method}            & \multirow{2}{*}{\begin{tabular}[c]{@{}c@{}}Training\\  Dataset\end{tabular}} & \multirow{2}{*}{\begin{tabular}[c]{@{}c@{}}Trainable\\ Parameters (M)\end{tabular}} & \multirow{2}{*}{Flops (G)} & \multicolumn{2}{c}{VoxCeleb-O} \\ \cline{5-6} 
                                   &                                                                              &                                                                                     &                            & EER           & minDCF         \\ \hline
ResNet34 (256)~\cite{chung2019voxsrc}             & VoxCeleb2~(Data Aug)                                                                    & 7.031                                                                               & 2.929                      & \textbf{1.42}          & -              \\
X-vector~\cite{wang2021revisiting}                           & VoxCeleb2~(Data Aug)                                                                    & 4.357                                                                               & 0.724                      & 2.41          & 0.260          \\
ECAPA-TDNN~\cite{desplanques2020ecapa}                           & VoxCeleb2~(Data Aug)                                                                    & 20.768                                                                               & 0.814                      & 0.87        &  0.107         \\
RawNet3~\cite{jung2022pushing}                           & VoxCeleb2 \& VoxCeleb1 (Data Aug)                                                                    &-                                                                               & -                      & 0.89          & 0.0659        \\
MFA-Conformer~\cite{zhang2022mfa}                            & VoxCeleb2 \& VoxCeleb1 \& SITW(Data Aug)                                                                    & 20.5                                                                              &  -                      & 0.64         & 0.081          \\
Whisper-SV                         & VoxCeleb2                                                                    & 5.336                                                                               & 0.643                      & \textbf{1.71}          & \textbf{0.211 }        \\ \hline
\end{tabular}}
\end{table}
\subsection{Ablation and Visualization}

\begin{figure}[th]
\centering    \centerline{\includegraphics[width=\linewidth]{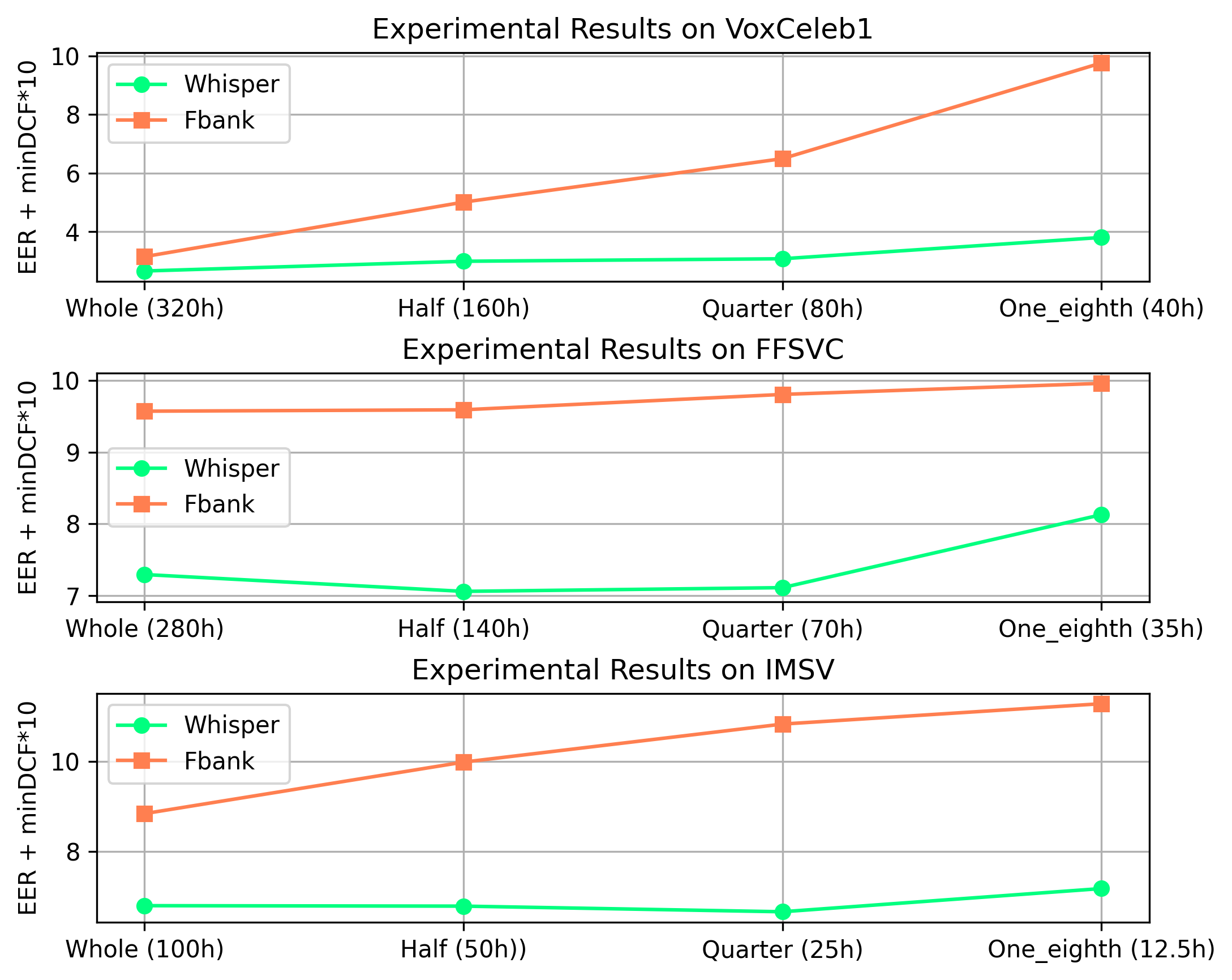}}
\caption{Experimental results of ECAPA-TDNN and Whisper-SV trained with different proportions of training data (The training duration (hours) corresponding to the value in parentheses on the x-axis).}
\label{fig:dataset_abliation}
\end{figure}

\textbf{Data Reduction Ablation}
To further investigate whether Whisper-SV performs better than Fbank in low-resource scenarios, we explore the correlation between the reduction in data and the performance of SV. The experimental results are illustrated in Fig. \ref{fig:dataset_abliation}. This analysis is carried out after partitioning the training data into various fractions, including the whole dataset, half, quarter, and one-eighth. The results presented in Fig. \ref{fig:dataset_abliation} show that the VoxCeleb1 dataset exhibits superior performance when utilizing the whole dataset. However, as the amount of data decreases, the performance degradation of ECAPA-TDNN trained with Fbank features is significantly more pronounced than Whisper-SV's results.

Meanwhile, experimental results from the FFSVC and IMSV datasets indicate that Whisper-SV achieves the lowest EER and minDCF when only a quarter of the data volume is used. In contrast, ECAPA-TDNN trained with Fbank achieves optimal results when the whole dataset is employed. These results suggest that Whisper representations contain a higher density of speaker-specific features while effectively filtering out extraneous noise elements. On the other hand, an excessive volume of speech data per speaker may lead to overfitting, making Whisper representations more advantageous for SV tasks in low-resource data scenarios. 

\begin{figure}[th]
\centering
    \centerline{\includegraphics[width=\linewidth]{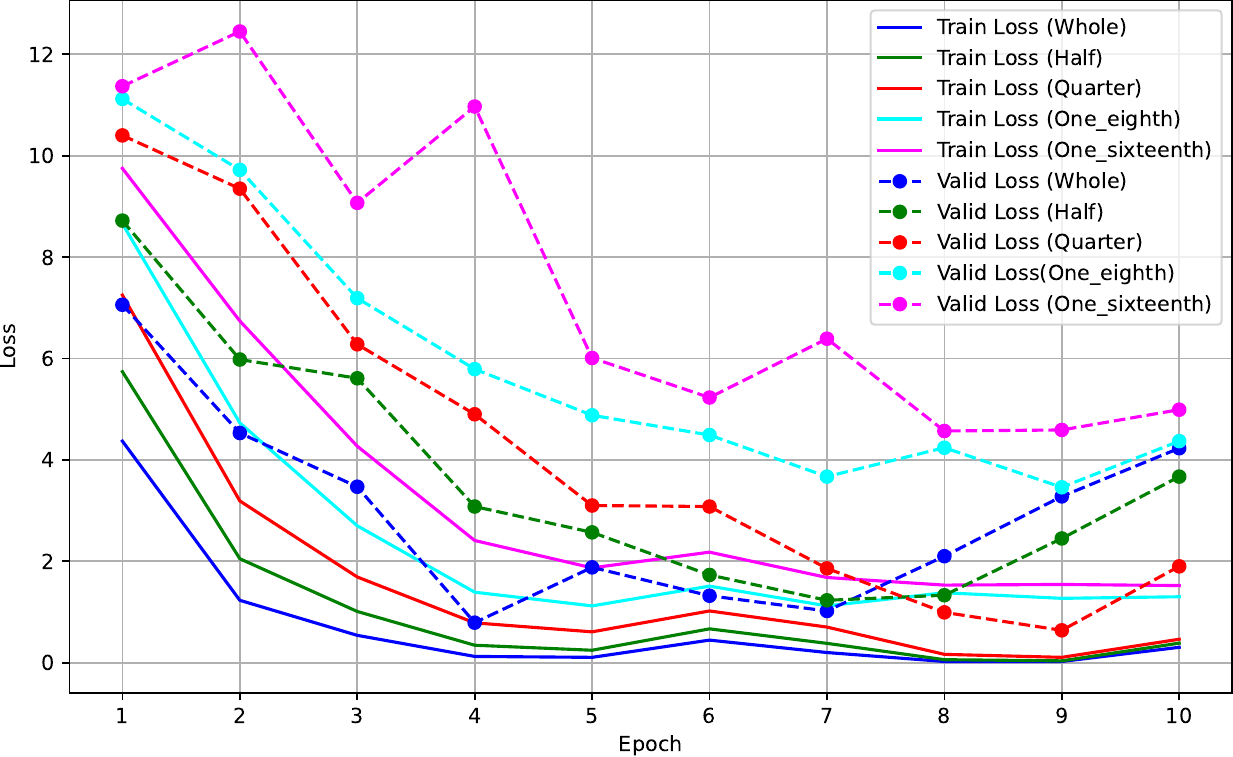}}
\caption{The training and validation loss value in the training process on FFSVC.  }
\label{fig:loss}
\end{figure}

The training and validation losses are depicted in Fig.~\ref{fig:loss}. 
Fig.~\ref{fig:loss} indicates that during the FFSVC training process, training with whole and half of the data shows signs of overfitting starting at epochs 7 and 8, as evidenced by increasing validation loss. In contrast, training with 1/8 and 1/16 of the data exhibits underfitting, whereas training with 1/4 of the data mitigates overfitting without leading to underfitting.

 
\textbf{Zoom Layer Channel Ablation}
Subsequently, we examine the influence of varying the scale of whisper representations within the zoom layer on the efficacy of Whisper-SV. The experimental results presented in TABLE \ref{channels_results} reveal that Whisper-SV attains optimal performance across three distinct datasets when the whisper representation is scaled to a dimensionality 128. Meanwhile, Whisper-SV with 128 channels in the zoom layer demonstrates fewer learnable parameters and better performance. This indicates that Whisper-SV is distinguished by its low computational complexity and minimal trainable parameters.

\begin{table}[th]
\caption{Experimental results on different channels in the zoom layer (EER(\%), minDCF (p=0.01)).}
\label{channels_results}
\resizebox{\linewidth}{!}{
\begin{tabular}{cccccccc}
\toprule
\hline
\multirow{2}{*}{Zoom layer Channel} & \multirow{2}{*}{\shortstack{Trainable \\ Parameters~(M)}} & \multicolumn{2}{c}{Voxceleb1} & \multicolumn{2}{c}{IMSV} & \multicolumn{2}{c}{FFSVC} \\ \cline{3-8}  
& & EER & minDCF & EER & minDCF & EER & minDCF \\ \hline
512 & 19.22 & 2.67 & 0.347 & 7.77 & 0.572 & 8.37 & 0.768 \\
256 & 9.845 & 2.54 & 0.365 & 7.33 & 0.585 & 8.10 & 0.789 \\
128 & 5.335 & \textbf{2.22} & \textbf{0.307} & \textbf{7.50} & \textbf{0.582} & \textbf{7.90} & \textbf{0.631} \\
64 & 3.586 & 3.63 & 0.398 & 8.55 & 0.612 & 9.43 & 0.803 \\ \hline
\end{tabular}}
 
\end{table}

\textbf{Channel Ablation in Multi-layer Aggregation Module }
The multi-layer aggregation module, consisting of 1D convolutional blocks, aggregates multi-layer representations extracted from Whisper. The experimental results of different channels in the multi-layer aggregation module are illustrated in TABLE \ref{mla_results}. The experimental results in TABLE \ref{mla_results} indicate that Whisper-SV exhibits optimal performance when the channel count of the multi-layer aggregation module is configured to 512.

\begin{table}[th]
\caption{Comparison of experimental results on different channels in the multi-layer aggregation module (EER(\%), minDCF (p=0.01)).}
\label{mla_results}
\resizebox{\linewidth}{!}{
\begin{tabular}{cccccccc}
\toprule
\hline
\multirow{2}{*}{\parbox{5cm}{ \centering Channels in Multi-layer \\ Aggregation Module}} & \multirow{2}{*}{\parbox{2cm}{Trainable \\ Parameters~(M)}} &  \multicolumn{2}{c}{Voxceleb1} & \multicolumn{2}{c}{IMSV} & \multicolumn{2}{c}{FFSVC} \\ \cline{3-8} 
& & EER & minDCF & EER & minDCF & EER & minDCF \\ \hline
512 & 5.335 & \textbf{2.22} & \textbf{0.307} & 7.50 & \textbf{0.582} & \textbf{7.90} & \textbf{0.631} \\
256 & 4.251 & 2.22 & 0.317 & \textbf{7.18} & 0.601 & 8.10 & 0.789 \\
128 & 3.774 & 2.51 & 0.351 & 7.974 & 0.67 & 8.79 & 0.814 \\ \hline
\end{tabular}}
\end{table}

\begin{figure*}[th]
\centering
\captionsetup{justification=centering} 
\centerline{\includegraphics[width=\linewidth]{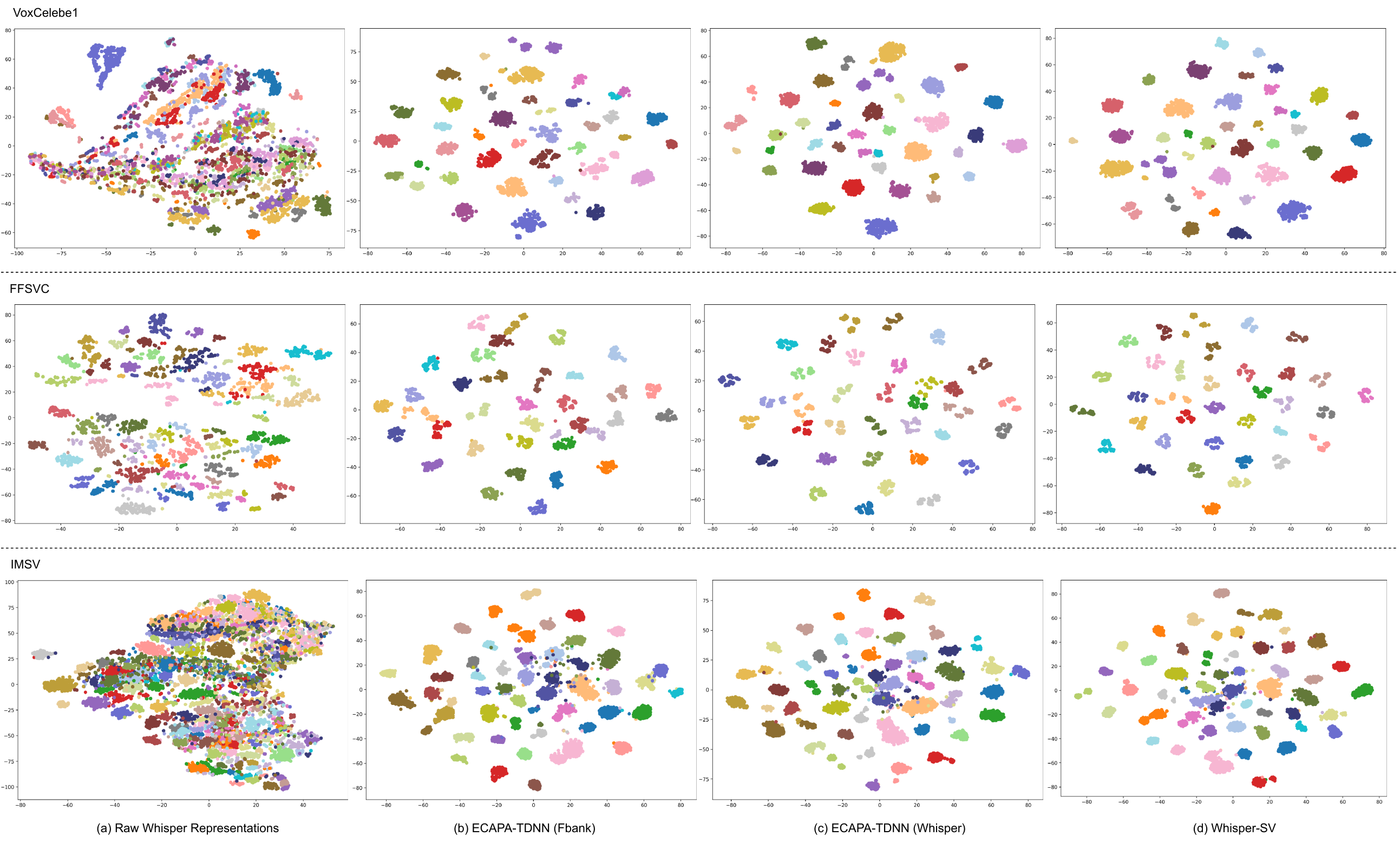}}
\caption{Embedding visualization of different SV models.}
\label{fig:whisper-sv_visualization}
\end{figure*}
\begin{figure}[th]
\centering
\captionsetup{justification=centering} 
\centerline{\includegraphics[width=\linewidth]{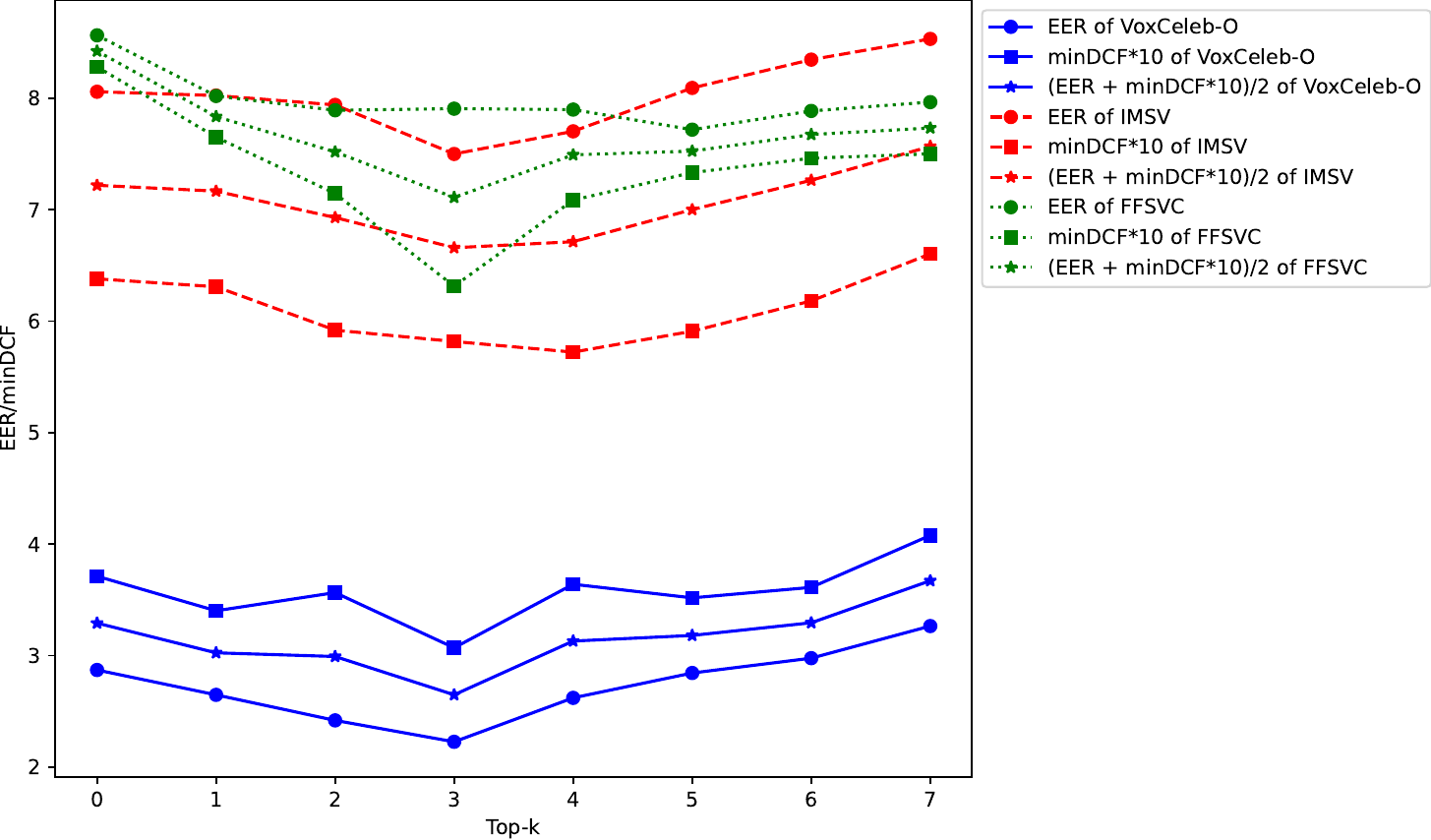}}
\caption{Experimental results on different top-k on VoxCeleb1~/~FFSVC~/~IMSV.}
\label{fig:TOP_K}
\end{figure}

\textbf{Different Top-k Ablation }
Fig. \ref{fig:TOP_K} displays the results of the ablation study conducted with different top-k values. Specifically, k is from the set of \{1, 2, 3, 4, 5, 6, 7, 8\}. The experimental results indicate that Whisper-SV achieves its best performance when k=4, signifying that selecting the most discriminative representations from the top four layers yields the best results. Interestingly, aggregating an excessive number of layers in Whisper representations does not reduce EER and minDCF. This lack of improvement may be attributed to the inclusion of non-speaker-related features introduced by the extensive layers of Whisper representations. Furthermore, the excessive aggregation of representations may lead to model overfitting.

\textbf{Embedding Visualization}
To further elucidate the discrimination of speaker embeddings extracted from different SV models, we employ T-SNE~\cite{van2008visualizing} to illustrate the raw whisper representations, embeddings generated by the ECAPA-TDNN (Fbank), embeddings generated by ECAPA-TDNN (Whisper), and embeddings generated by Whisper-SV models. The distribution of the above embeddings is depicted in Fig.~\ref{fig:whisper-sv_visualization}. In Fig.~\ref{fig:whisper-sv_visualization}, we can see that irrespective of the testing dataset, utilizing raw whisper representations for speaker identification is 
infeasible. The entanglement of speaker embeddings demonstrates we need an adaptor to transfer Whisper to SV tasks.
Observing the embedding distribution on VoxCeleb1, it becomes apparent that ECAPA-TDNN (Whisper) exhibits enhanced intra-class cohesion compared to ECAPA-TDNN (Fbank). Furthermore, Whisper-SV demonstrates improved cohesion in its embedding distribution and greater inter-class separation. These observations hold when analyzing the results on FFSVC, further substantiating that speaker embeddings of Whisper-SV are more discriminative and robust. The visualization results for Whisper-SV on IMSV show an improvement in classification performance on hard samples when contrasted with ECAPA-TDNN (Fbank) and ECAPA-TDNN (Whisper).

\section{Conclusion}
\label{sec:typestyle}
This paper proposes an adaptor framework named Whisper-SV, designed to adapt Whisper for SV tasks, particularly benefiting low-data-resource SV scenarios. Whisper-SV consists of four distinct modules: a pre-trained Whisper module, a representation selection module, a multi-layer aggregation module, and a speaker classifier module. Specifically, the pre-trained Whisper module is the robust and generalized representation extractor. As Whisper is not explicitly tailored for SV tasks, the representation selection module is introduced to assess speaker identity characteristics within each layer of Whisper. It selects the top-k layers with prominent speaker-related features. Subsequently, the multi-layer aggregation module is designed to fuse representations from the selected top-k layers to a compacted representation abundant in speaker-specific discriminative characteristics. Finally, the classifier module is used for speaker classification. Attributable to Whisper's pre-training on massive and diverse datasets, Whisper-SV achieves remarkable performance with a modest number of trainable parameters and a limited amount of training speech. This demonstrates the suitability of Whisper-SV for low-data-resource SV tasks. In addition, experimental results and analysis illustrate the superior performance of Whisper-SV compared to other competitive methods in low-data-resource SV tasks. 

\section{Future Work}
Whisper-SV achieves superior performance in low-data-resource SV tasks. However, since Whisper is a parameter-heavy, large speech foundation model, it is not well-suited for SV applications requiring low computational resources and fast inference speeds. Therefore, in future work, we will focus on developing SV models with Whisper using transfer learning methods to reduce reliance on Whisper during inference.

\bibliographystyle{IEEEbib}
\bibliography{refs}
\end{document}